\documentstyle[aps,twocolumn]{revtex}
\begin{document}
\author{LiXiang Cen$^1,2$, XinQi Li$^1,2$, YiJing Yan$^2$, HouZhi Zheng$^1$, and ShunJin
Wang$^3$}
\address{$^{1}$NLSM, Institute of Semiconductors, The Chinese Academy of Sciences, 
Beijing 100083, People's Republic of China\\
$^{2}$Department of Chemistry, Hong Kong University of Science and Technology, 
Kowloon, Hong Kong\\
$^{3}$Department of Physics, Sichuan University, Chengdu 610064, People's Republic 
of China}
\title{Evaluating holonomic quantum computation: beyond adiabatic limitation}
\maketitle

\begin{abstract}
The proposal of the optical scheme for holonomic quantum computation is
evaluated based on dynamical resolution to the system beyond adiabatic
limitation. The time-dependent Schr\"{o}dinger equation is exactly solved by
virtue of the cranking representation and gauge transformation approach.
Besides providing rigorous confirmation to holonomies of the geometrical
prediction that holds for the ideally adiabatic situation, the dynamical
resolution enables one to evaluate elaborately the amplitude of the
nonadiabatic deviation, so that the errors induced to the quantum
computation can be explicitly estimated.
\end{abstract}
\pacs{PACS numbers: 03.67.Lx, 03.65.Vf}

The recently proposed holonomic approach to quantum computation \cite{zanar}%
- \cite{hqc} surely predicts a striking contribution to the application of
quantum physics. Transcending the traditional dynamical means for quantum
computation, the holonomic approach realizes quantum information processing
by endowing the quantum code with a non-trivial global topology (a gauge
field potential) and the associated holonomies then allow for the universal
quantum computing. Specifically, in the scheme of holonomic quantum
computation, information is encoded in a degenerate eigenspace of the
governing Hamiltonian and the holonomies (abelian as well as non-abelian) 
\cite{berry}-\cite{wilzek} are acquired by driving the system to undergo
appropriate loops in the parameter space adiabatically. Besides suggesting
an intriguing connection between the gauge fields and the information
processing, such a geometrical means for quantum manipulation is believed to
have built-in fault-tolerant features \cite{fault}-\cite{falci} due to its
inherent stability against local perturbations. Considerable attention has
been addressed to this topic recently and the all-geometrical implementation
for universal quantum gates has been proposed by optical schemes, based on
laser manipulation of ions confined in a Paul trap \cite{duan} or neutral
atoms in an optical resonator \cite{new}.

The existent exploration for holonomic quantum computation is based on the
analysis by pure geometrical fashion. It is true that in the adiabatic limit
the holonomy associated with the evolving loop is determined by the path
traced by the time-evolution ray and the curvature of the ray space. This
involves the abelian holonomy (the Berry phase) and the non-abelian one
merely known as adiabatic connection. Nevertheless, as a whole physical
problem, as the dynamics of a system generates a time-dependent physical
state, a specified geometrical object (the ray) is generated as well. In
such a sense, dynamics determines the holonomy through determining the ray
itself and its path. Moreover, in view that the realistic evolution of a
physical system could not be ideally adiabatic and the nonadiabaticity shall
alter the time-evolution of the ray and thus inevitably induce deviation
from the adiabatic consequences. The evaluation of such deviation and the
resulting errors in quantum computation is definitely a dynamical problem
that goes beyond the geometrical exploration.

In this paper we employ a tractable model of the optical scheme to exploit
this subject. For the appropriately chosen loops of the Hamiltonian in the
parameter space, the time-dependent Schr\"{o}dinger equation is exactly
solved by virtue of the cranking representation and gauge transformation
approach. The derived dynamical evolution of the system recovers the
holonomic transformation provided by geometrical consequences, including the
simple abelian phase factor and the general non-abelian operation. Thus our
results provide further confirmation of the geometrical prediction, and
besides, the errors caused by nonadiabatic effects for the holonomic quantum
gate operation can be estimated explicitly.

For the proposed optical scheme of holonomy quantum computation \cite
{duan,new}, the basic idea relies on the adiabatic passage via the dark
states since the dynamical evolution restricted to such a space is
completely trivial. The system encoding the qubit is realized by a
four-level $\Lambda $-type trapped ion (or a similar cavity atom). The three
ground levels $|g_{i}\rangle $ ($i=1,2,3$) are highly degenerate and each
couples to the excited state $|e\rangle $ in a tunable way. The states $%
|g_{1}\rangle $ and $|g_{2}\rangle $ stand for the computational bases $%
|0\rangle $ and $|1\rangle $, respectively, and $|g_{3}\rangle $ is an
ancillary level required for implementation of gate operations. Such a
system admits two dark states that have no contribution from the excited
state. Through changing the Rabi frequencies and driving the dark states to
undergo appropriate cyclic evolutions in an adiabatic fashion, the universal
single-bit gate operations $e^{i\phi |1\rangle \langle 1|}$ and $e^{i\phi
\sigma _{y}}$ can be generated due to the global geometry of the bundle of
the eigenspace of the dark states.

To evaluate the gate operation $e^{i\phi |1\rangle \langle 1|}$ from a
dynamical viewpoint, let us explore the state evolution generated by the
periodic Hamiltonian \cite{duan,new} 
\begin{equation}
H(t)=\Omega \sin \theta (\sigma _{2e}+\sigma _{e2})+\Omega \cos \theta
(\sigma _{3e}e^{i\varphi }+\sigma _{e3}e^{-i\varphi }),  \label{ham1}
\end{equation}
where $\theta $ is a fixed parameter and $\varphi $ is assumed to rotate at
a constant frequency $\gamma $ for convenience. The equation of motion for
the system is 
\begin{equation}
i\frac{\partial }{\partial t}|\Psi (t)\rangle =H(t)|\Psi (t)\rangle .
\label{schro}
\end{equation}
It is known that in the adiabatic limit, the geometrical exploration shows
that the dark state of the system, $|D(t)\rangle =\cos \theta |g_{2}\rangle
-\sin \theta e^{i\gamma t}|g_{3}\rangle $, shall acquire a net Berry phase 
\cite{berry,simon} after a period $T=2\pi /\gamma $: $|D(T)\rangle =e^{i\phi
}|D(0)\rangle $ with $\phi =4\pi \sin ^{2}\theta $. To give a dynamical
resolution to the system beyond adiabatic limitation, we use the cranking
representation, that is, the Hamiltonian (\ref{ham1}) can be regarded as a
cranked one 
\begin{equation}
H(t)=e^{i\gamma t\sigma _{33}}H_{0}e^{-i\gamma t\sigma _{33}},  \label{crank}
\end{equation}
where 
\begin{equation}
H_{0}=\Omega \sin \theta (\sigma _{2e}+\sigma _{e2})+\Omega \cos \theta
(\sigma _{3e}+\sigma _{e3}),  \label{h0}
\end{equation}
and the unitary transformation $e^{i\gamma t\sigma _{33}}$ can be viewed as
an element of the SU(3) group. Consequently, the dynamical invariant of the
system can be shown as 
\begin{equation}
I(t)=e^{i\gamma t\sigma _{33}}(H_{0}+\gamma \sigma _{33})e^{-i\gamma t\sigma
_{33}}=H(t)+\gamma \sigma _{33},  \label{invar}
\end{equation}
which satisfies \cite{lewis} 
\begin{equation}
\frac{dI(t)}{dt}=\frac{\partial I(t)}{\partial t}-i[I(t),H(t)]=0.
\label{inveq}
\end{equation}
The second term of (\ref{invar}) accounts for an extra gauge potential since 
$H(t)$ depends on time explicitly. Now the recurrent basis $|\psi (t)\rangle 
$ of the system [it differs from the basic solution $|\Psi (t)\rangle $ of
the Schr\"{o}dinger equation (\ref{schro}) only by a phase factor] can be
obtained by solving the instantaneous eigensolutions of $I(t)$. It turns
out, one only needs to solve the characteristic equation 
\begin{equation}
x^{3}-(\gamma /\Omega )x^{2}-x+(\gamma /\Omega )\sin ^{2}\theta =0
\label{chareq}
\end{equation}
and the eigenvalues of $I(t)$ are given by $E_{i}(\frac{\gamma }{\Omega }%
)=\Omega x_{i}(\frac{\gamma }{\Omega })$ ($i=-1,0,1$). It is straightforward
to show that the recurrent basis $|\psi _{0}(t)\rangle $ represented by the
middle number $E_{0}$ approaches to the dark state $|D(t)\rangle $ in the
adiabatic limit $\gamma /\Omega \rightarrow 0$. Now the leakage error
induced by the nonadiabatic effect can be conveniently estimated by the
overlap [see Fig. 1(a)] 
\begin{equation}
\eta (\theta ,\frac{\gamma }{\Omega })=|\langle \psi _{0}(0)|D(0)\rangle
|^{2}=|\langle \psi _{0}(T)|D(T)\rangle |^{2}.  \label{overlap}
\end{equation}

Besides the leakage, the nonadiabatic evolution shall result in deviation to
the desired phase factor. It follows, instead of the net Berry phase, the
cyclic evolution here induces a total phase (the so-called Lewis-Riesenfeld
phase) 
\begin{equation}
\Phi =\int_{0}^{T}\langle \psi _{0}(t)|i\frac{\partial }{\partial t}%
-H(t)|\psi _{0}(t)\rangle dt=E_{0}\frac{2\pi }{\gamma }.  \label{phase}
\end{equation}
The detailed depiction of the deviation for the phase factor is shown in
Fig. 1(b). Noting that in the adiabatic limit, the total phase \cite{note} 
\begin{equation}
\Phi =\lim_{\gamma /\Omega \rightarrow 0}2\pi \frac{x_{0}(\frac{\gamma }{%
\Omega })}{\gamma /\Omega }=4\pi \sin ^{2}\theta ,  \label{limit}
\end{equation}
the geometrical consequence is thus recovered.

The validity of the above evaluation is based on a presumption that the
initial state $|D(0)\rangle =\cos \theta |g_{2}\rangle -\sin \theta
|g_{3}\rangle $ can be generated from the computational basis $|g_{2}\rangle 
$ and so the inverse process. Explicitly, such processes can be accomplished
by the driven Hamiltonian (\ref{ham1}) through changing the parameter $%
\theta $ adiabatically. Conventionally, the nonadiabatic effect here shall
lead to an additional error for the quantum computation. However, such an
error can be in principle avoided through appending a matching interaction
to compensate the gauge potential term induced to the system. Specifically,
one can use the following Hamiltonian (setting $\varphi =0$) 
\begin{equation}
H_{tot}(t)=H(t)+H_{ad}(t),H_{ad}(t)=i\dot{\theta}(t)(\sigma _{23}-\sigma
_{32}).  \label{addh}
\end{equation}
It follows that the dynamical invariant of the system $H_{tot}(t)$ now has a
form $I(t)=H(t)$, thus the above state transformation can be processed
exactly. Physically, the interaction $H_{ad}(t)$ can be realized by a
microwave coupling to the two degenerate levels $|g_{2}\rangle $ and $%
|g_{3}\rangle $, with its intensity accurately controlled through a
derivative feedback process.

Now we investigate the gate operation $e^{i\phi \sigma _{y}}$ achieved by
the holonomic means. The corresponding evolution is generated by the
Hamiltonian 
\begin{eqnarray}
H(t) &=&\Omega \sin \theta \cos \varphi (\sigma _{1e}+\sigma _{e1}) 
\nonumber \\
&+&\Omega \sin \theta \sin \varphi (\sigma _{2e}+\sigma _{e2})+\Omega \cos
\theta (\sigma _{3e}+\sigma _{e3}),  \label{ham2}
\end{eqnarray}
where the parameter $\varphi =\gamma t$. As is known, the adiabatic cyclic
evolution of the Hamiltonian generates a non-abelian holonomy due to its
degeneracy structure of the dark states. It can be easily worked out, from
the formula of Ref. \cite{wilzek}, that the holonomic transformation 
\begin{equation}
u_{C}=e^{i2\pi \cos \theta D_{y}},  \label{holon}
\end{equation}
where $D_{y}=i(|D_{2}\rangle \langle D_{1}|-|D_{1}\rangle \langle D_{2}|)$,
and the two dark states, $|D_{1}\rangle =\cos \theta |g_{1}\rangle -\sin
\theta |g_{3}\rangle $ and $|D_{2}\rangle =|g_{2}\rangle $, span the
degenerate space of the starting (ending) Hamiltonian. Note that the
Hamiltonian (\ref{ham2}) possesses an su(4) Lie algebraic structure and
dynamical resolution to the system is usually very complicated.
Surprisingly, as we shall show in the following, this system can be exactly
solved by the gauge transformation approach \cite{gauge,gaug2}, and its
dynamical evolution analytically manifested thus leads to a complete
understanding of the adiabatic and nonadiabatic properties for the
time-dependent Hamiltonian system.

Similar to the cranking method used above, we introduce the unitary gauge
transformation 
\begin{equation}
U_{g}(t)=e^{-\gamma t(\sigma _{12}-\sigma _{21})}  \label{gauge2}
\end{equation}
to the equation of motion for the system, from which a covariant
Schr\"{o}dinger equation is stemmed 
\begin{eqnarray}
|\Psi _{g}(t)\rangle &=&U_{g}^{-1}(t)|\Psi (t)\rangle ,  \nonumber \\
i\frac{\partial }{\partial t}|\Psi _{g}(t)\rangle &=&H_{g}|\Psi
_{g}(t)\rangle ,  \label{schro2}
\end{eqnarray}
with the gauged Hamiltonian 
\begin{eqnarray}
H_{g} &=&U_{g}^{-1}HU_{g}-iU_{g}^{-1}\frac{\partial U_{g}}{\partial t} 
\nonumber \\
&=&\Omega \sin \theta (\sigma _{1e}+\sigma _{e1})  \nonumber \\
&+&\Omega \cos \theta (\sigma _{3e}+\sigma _{e3})+i\gamma (\sigma
_{12}-\sigma _{21}).  \label{gaugeh}
\end{eqnarray}
In view that the above Hamiltonian is time independent, the basic solutions $%
|\Psi _{g}^{n}(t)\rangle $ to the covariant equation (\ref{schro2}) can be
easily obtained and the corresponding eigenvalues are as follows 
\begin{eqnarray}
E_{1,2} &=&\pm \frac{\sqrt{2}}{2}\bar{\Omega}\left[ 1-\sqrt{1-4(\frac{\gamma 
}{\Omega })^{2}\cos ^{2}\bar{\theta}}\right] ^{1/2},  \nonumber \\
E_{3,4} &=&\pm \frac{\sqrt{2}}{2}\bar{\Omega}\left[ 1+\sqrt{1-4(\frac{\gamma 
}{\Omega })^{2}\cos ^{2}\bar{\theta}}\right] ^{1/2},  \label{eigenv}
\end{eqnarray}
where 
\begin{equation}
\bar{\Omega}=\Omega \sqrt{1+(\gamma /\Omega )^{2}},\ \ \ \cos \bar{\theta}=%
\frac{\cos \theta }{1+(\gamma /\Omega )^{2}}.  \label{param}
\end{equation}
The dynamical basis of the system (\ref{ham2}) can be directly obtained as $%
|\Psi _{n}(t)\rangle =U_{g}|\Psi _{g}^{n}(t)\rangle $, from which one can
see that $E_{n}$ has the natural implication related to the total phase. Now
the time evolution operator generated by the Hamiltonian (\ref{ham2}) can be
given 
\begin{eqnarray}
U_{C}(T) &=&\sum_{n=1}^{4}|\Psi _{n}(T)\rangle \langle \Psi _{n}(0)| 
\nonumber \\
&=&\sum_{n=1}^{4}e^{-iE_{n}\frac{2\pi }{\gamma }}|\Psi _{n}(0)\rangle
\langle \Psi _{n}(0)|.  \label{timeope}
\end{eqnarray}
Considering the asymptotic behavior of the evolution in the adiabatic limit,
it follows that $\lim_{\gamma /\Omega \rightarrow 0}\frac{E_{1,2}}{\gamma }%
=\pm \cos \theta $, and the phase-equipped dynamical bases $|\Psi
_{1}(t)\rangle $ and $|\Psi _{2}(t)\rangle $ have the form 
\begin{eqnarray}
|\Psi _{1}(t)\rangle &=&\frac{\sqrt{2}}{2}e^{-i\gamma t\cos \theta }[(\cos
\theta \cos \gamma t+i\sin \gamma t)|g_{1}\rangle  \nonumber \\
&&+(\cos \theta \sin \gamma t-i\cos \gamma t)|g_{2}\rangle -\sin \theta
|g_{3}\rangle ],  \nonumber \\
|\Psi _{2}(t)\rangle &=&\frac{\sqrt{2}}{2}e^{i\gamma t\cos \theta }[(\cos
\theta \cos \gamma t-i\sin \gamma t)|g_{1}\rangle  \nonumber \\
&&+(\cos \theta \sin \gamma t-i\cos \gamma t)|g_{2}\rangle -\sin \theta
|g_{3}\rangle ].  \label{dark2}
\end{eqnarray}
One can verify that they are the instantaneous eigenstates of the
Hamiltonian (\ref{ham2}) with a two-degeneracy eigenvalue $0$, and the
equipped phases are just the Berry phases accordingly. Thus the cyclic
evolution restricted to the space spanned by these two states is purely
geometrical and can be denoted as 
\begin{equation}
u(T)=e^{-i2\pi \cos \theta }|\Psi _{1}(0)\rangle \langle \Psi
_{1}(0)|+e^{i2\pi \cos \theta }|\Psi _{2}(0)\rangle \langle \Psi _{2}(0)|
\label{dyholo}
\end{equation}
with $|\Psi _{1}(0)\rangle =(|D_{1}\rangle -i|D_{2}\rangle )/\sqrt{2}$ and $%
|\Psi _{2}(0)\rangle =(|D_{1}\rangle +i|D_{2}\rangle )/\sqrt{2}$. It can be
easily recognized that the operator (\ref{dyholo}) is just the non-abelian
holonomy (\ref{holon}), thus the geometrical nature is verified again.

The above dynamical resolution to the system is important. Besides offering
a vivid verification to the remarkable formula of non-abelian holonomy \cite
{wilzek}, which holds for the ideally adiabatic situation, it enables one to
evaluate elaborately the amplitude of the nonadiabaticity deviation and the
resulting errors to the holonomic gate operation $e^{i\phi \sigma _{y}}$. In
detail, the population transfer from the initial state $|\Psi (0)\rangle
=|g_{2}\rangle $ is pictured in Fig. 2. The leakage out of the computational
space can be estimated by the projection (see also Fig. 2) 
\begin{equation}
\eta (\theta ,\frac{\gamma }{\Omega })=\sum_{i=1}^{2}|\langle
D_{i}|U(T)|\Psi (0)\rangle |^{2}.  \label{proj2}
\end{equation}

Similar to the former case, to transform the computational basis $%
|g_{1}\rangle $ into the dark state $|D_{1}\rangle =\cos \theta
|g_{1}\rangle -\sin \theta |g_{3}\rangle $ and to invert the process
successfully, one needs to use the Hamiltonian (\ref{ham2}) (with $\theta $
tunable and $\varphi =0$) along with a matching interaction 
\begin{equation}
H_{tot}(t)=H(t)+H_{ad}(t),H_{ad}(t)=i\dot{\theta}(t)(\sigma _{13}-\sigma
_{31}).  \label{addh2}
\end{equation}

Up to now, we have investigated the single-qubit holonomic operations of the
optical scheme and evaluated the nonadiabaticity-causing errors for quantum
computation. It deserves to point out that, the proposal \cite{duan} of
geometrical implementation for the controlled two-qubit phase shift gate, $%
e^{i\phi |11\rangle \langle 11|}$, which is sufficient for the universal
quantum computation along with the two single-qubit gates, can be explored
in a similar way. The scheme is realized by two-color laser manipulation on
the trapped ions \cite{ions}. Briefly, the transition from the ground states 
$|g_{2}\rangle $ and $|g_{3}\rangle $ to the excited state $|e\rangle $ is
driven by two different bi-chromatic laser beams with their amplitudes and
phases of the Rabi frequencies controllable. The frequencies of the laser
fields are tuned so that the two-photon process, exciting pair ions, is
resonant and the single-photon excitation is off-resonant. Hence the system
can be described by an effective Hamiltonian in the notation of Ref. \cite
{duan} 
\begin{eqnarray}
H_{eff} &\propto &-|\Omega _{1}|^{2}(e^{i2\varphi _{1}}|ee\rangle \langle
g_{2}g_{2}|+h.c.)  \nonumber \\
&+&|\Omega _{2}|^{2}(e^{i2\varphi _{2}}|ee\rangle \langle g_{3}g_{3}|+h.c.),
\label{ham3}
\end{eqnarray}
where the relative intensity of the Rabi frequencies $\tan \theta =-|\Omega
_{1}|^{2}/|\Omega _{2}|^{2}$ and the phase difference $\varphi /2=\varphi
_{1}-\varphi _{2}$ are tunable. One can see that the bases $%
|g_{1}g_{1}\rangle $, $|g_{1}g_{2}\rangle $, and $|g_{2}g_{1}\rangle $ are
decoupled from the evolution, and the component $|g_{2}g_{2}\rangle $
serving as the code $|11\rangle $ evolved in an enclosed space spanned by \{$%
|g_{2}g_{2}\rangle ,|g_{3}g_{3}\rangle ,|ee\rangle $\}. Introducing the
su(3) generators explicitly 
\begin{eqnarray}
A_{e2} &=&e^{i2\varphi _{1}}|ee\rangle \langle
g_{2}g_{2}|,~A_{e3}=e^{i2\varphi _{1}}|ee\rangle \langle g_{3}g_{3}|, 
\nonumber \\
A_{23} &=&|g_{2}g_{2}\rangle \langle g_{3}g_{3}|,~~~~~A_{\mu \nu }^{\dagger
}=A_{\nu \mu },  \label{operator}
\end{eqnarray}
the Hamiltonian (\ref{ham3}) can be rewritten as 
\begin{equation}
H_{eff}=g\sin \theta (A_{2e}+A_{e2})+g\cos \theta (A_{3e}e^{i\varphi
}+A_{e3}e^{-i\varphi }).  \label{ham33}
\end{equation}
Obviously this Hamiltonian possesses an su(3) algebraic structure isomorphic
to that of system (\ref{ham1}), thus all the discussions therein also hold
for the present system.

It should be noted that, the effective Hamiltonian (\ref{ham3}), respecting
a second-order process of the interaction, is quite a rough description of
the model. Specifically, it ignores the same second-order process induced by
virtual photons excitation in the self-transitions of the states: $%
|g_{2(3)}\rangle \rightarrow |g_{2(3)}\rangle $ and $|e\rangle \rightarrow
|e\rangle $. It can be anticipated that such self transitions shall dress
the energy levels of the ions and lift the degeneracy of the ground states,
which in turn affects the desired gate operation. Detailed exploration of
this point shall be presented in a future report.

This work was supported in part by the Postdoctoral Science Foundation, the
special funds for Major State Basic Research Project No. G001CB3095, the
National Natural Science Foundation No. 10175029 of China, and the Research
Grants Council of the Hong Kong Government.

{\bf Caption of Fig. 1:}

Deviation induced by nonadiabaticity for abelian holonomy. (a) The overlap $%
\eta $ for the parameters $\theta \in (0,\pi /2)$ and $\gamma /\Omega \in
[0,1]$; (b) The total phase $\Phi $ for $\theta \in [0,\pi ]$ and $\gamma
/\Omega \in [0,1]$.

{\bf Caption of Fig. 2:}

Deviation induced by nonadiabaticity for non-abelian holonomy. The initial
state is prepared in $|D_{2}\rangle $. The two solid curves show the results
for the population of the target state on $|D_{1}\rangle $ and $%
|D_{2}\rangle $, as a function of $1-\cos \theta $, respectively. The dashed
curve depicts the total population $\eta $ on the computational space.
Figures (a), (b), (c) and (d) correspond to $\gamma /\Omega =0.01,0.2,0.5$
and $0.8$, respectively.


\begin{references}
\bibitem{zanar}  P. Zanardi and M. Rasetti, Phys. Lett. A {\bf 264}, 94
(1999).

\bibitem{pach}  J. Pachos, P. Zanardi and M. Rasetti, Phys. Rev. A {\bf 61},
010305(R) (2000).

\bibitem{pach2}  J. Pachos and S. Chountasis, Phys. Rev. A {\bf 62}, 052318
(2000).

\bibitem{hqc}  J. Pachos and P. Zanardi, Intl. J. Mod. Phys. B {\bf 15},
1257 (2001).

\bibitem{berry}  M.V. Berry, Proc. R. Soc. London, Ser. A {\bf 392}, 45
(1984).

\bibitem{simon}  B. Simon, Phys. Rev. Lett. {\bf 51}, 2167 (1983).

\bibitem{wilzek}  F. Wilczek and A. Zee, Phys. Rev. Lett. {\bf 52}, 2111
(1984).

\bibitem{fault}  J. Preskill, {\it Fault-tolerant quantum computation} in 
{\it Introduction to quantum computaion and information}, Hoi-Kwong Lo, S.
Popescu and T. Spiller Eds., Wold Scientific, Singapore, 1999.

\bibitem{jone}  J.A. Jones V.Vedral, A. Ekert and G. Castagnoli, Nature
(London) {\bf 403}, 869 (2000).

\bibitem{ekert}  A. Ekert et al., J. Mod. Opt. {\bf 47}, 2501 (2000).

\bibitem{falci}  G. Falci, R. Fazio, G.M. Palma, J. Siewert and V.Vedral,
Nature (London) {\bf 407}, 355 (2000).

\bibitem{duan}  L.M. Duan, J.I. Cirac and P. Zoller, Science {\bf 292}, 1695
(2001).

\bibitem{new}  A. Recati, T. Calarco, P. Zanardi, J.I. Cirac and P. Zoller,
LANL e-print quant-ph/0204030.

\bibitem{lewis}  H.R. Lewis Jr., Phys. Rev. Lett. {\bf 18}, 510 (1967); H.R.
Lewis Jr. and W.B. Riesenfeld, J. Math. Phys. {\bf 10}, 1458 (1969).

\bibitem{note}  The limitation of Eq. (\ref{limit}) can be directly worked
out from the derivative on the hidden function $x_{0}=x_{0}(\frac{\gamma }{%
\Omega })$ determined by Eq. (\ref{chareq}).

\bibitem{gauge}  S.J. Wang, F.L. Li and A. Weiguny, Phys. Lett. A {\bf 180},
189 (1993).

\bibitem{gaug2}  S. Wang, B. Hu, Q. Jie, and B. Li, {\it Algebraic dynamics,
gauge transformation, and geometric phase} (private communication).

\bibitem{ions}  K, Molmer and A. Sorensen, Phy. Rev. Lett. {\bf 82}, 1835
(1999).
\end{references}
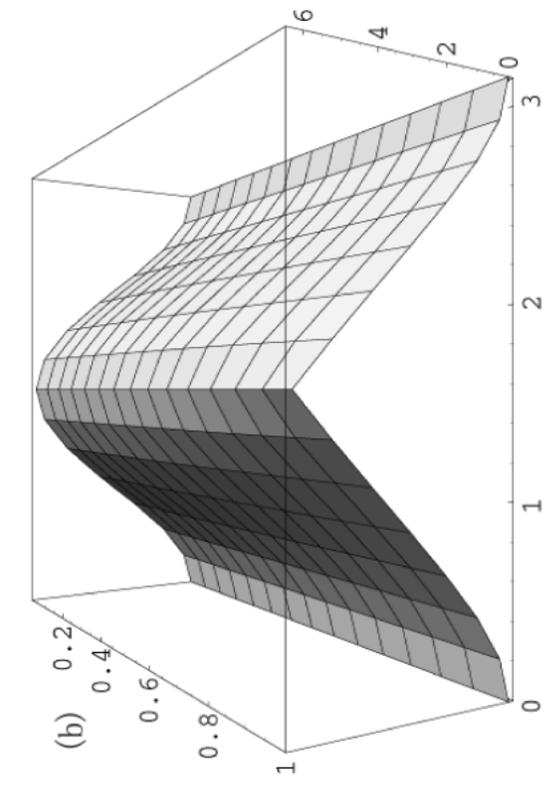
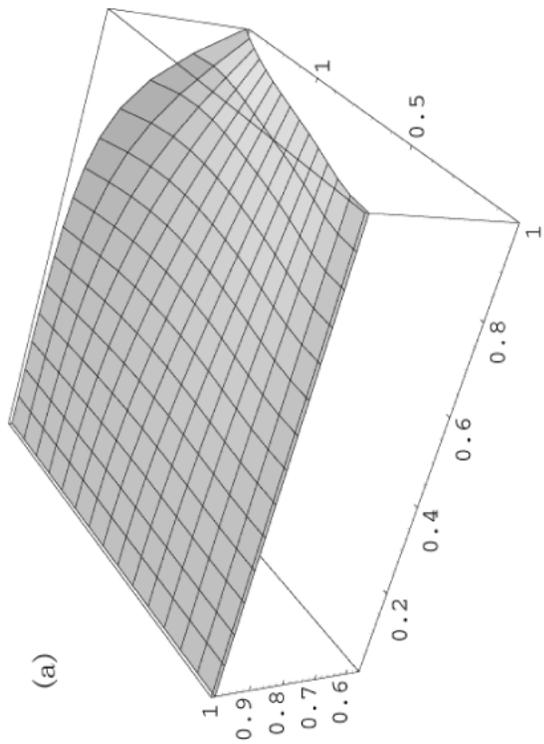
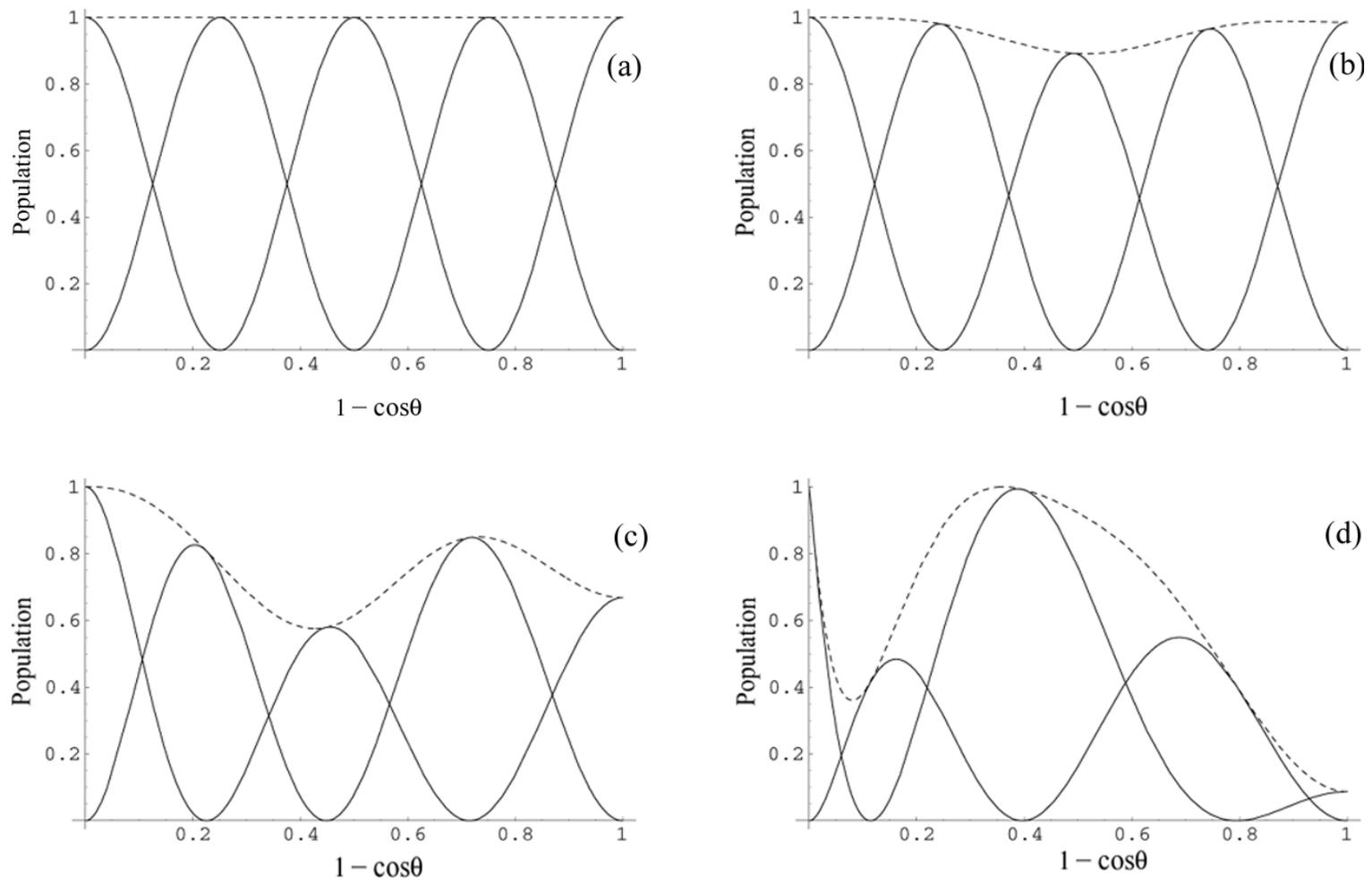
\end{document}